\documentclass[sigconf]{acmart}

\usepackage{booktabs} 
\usepackage{upgreek} 
\usepackage{tikz-cd}
\usepackage{tikz}
\usetikzlibrary{shapes,arrows}
\usepackage{hyperref}
\usepackage{graphicx,verbatimbox}
\usepackage{listings}
\usepackage{subcaption}
\usepackage{mathtools}
\usepackage{color}
\usepackage{tabularx,ragged2e}
\usepackage{multirow}
\usepackage{mathtools}
\usepackage{enumitem}
\usepackage[british]{babel}
\usepackage{xspace} 

\copyrightyear{2021} 
\acmYear{2021} 
\setcopyright{acmlicensed}\acmConference[ADCS '21]{Australasian Document Computing Symposium}{December 9, 2021}{Virtual Event, Australia}
\acmBooktitle{Australasian Document Computing Symposium (ADCS '21), December 9, 2021, Virtual Event, Australia}
\acmPrice{15.00}
\acmDOI{10.1145/3503516.3503530}
\acmISBN{978-1-4503-9599-1/21/12}

\newcommand{\mt}{MeSH terms\xspace}
\newcommand{\mts}{MeSH term suggestion\xspace}

\begin{document}
\title{MeSH Term Suggestion for Systematic Review Literature Search}
\author{Shuai Wang}
\affiliation{%
	\institution{University of Queensland}
	\city{Brisbane}
	\country{Australia}
}
\email{shuai.wang2@uq.edu.au}

\author{Hang Li}
\affiliation{%
  \institution{University of Queensland}
  \city{Brisbane}
  \country{Australia}
}
\email{hang.li4@uq.net.au}

\author{Harrisen Scells}
\affiliation{%
  \institution{University of Queensland}
  \city{Brisbane}
  \country{Australia}
}

\email{uqhscell@uq.edu.au}
\author{Daniel Locke}
\affiliation{%
  \institution{University of Queensland}
  \city{Brisbane}
  \country{Australia}
}

\email{d.locke@uq.edu.au}
\author{Guido Zuccon}
\affiliation{%
  \institution{University of Queensland}
  \city{Brisbane}
  \country{Australia}
}
\email{g.zuccon@uq.edu.au}

\begin{abstract}
High-quality medical systematic reviews require comprehensive literature searches to ensure the recommendations and outcomes are sufficiently reliable. Indeed, searching for relevant medical literature is a key phase in constructing systematic reviews and often involves domain (medical researchers) and search (information specialists) experts in developing the search queries. Queries in this context are highly complex, based on Boolean logic, include free-text terms and index terms from standardised terminologies (e.g., MeSH), and are difficult and time-consuming to build. The use of MeSH terms, in particular, has been shown to improve the quality of the search results. However, identifying the correct MeSH terms to include in a query is difficult: information experts are often unfamiliar with the MeSH database and unsure about the appropriateness of MeSH terms for a query. Naturally, the full value of the MeSH terminology is often not fully exploited.

This paper investigates methods to suggest MeSH terms based on an initial Boolean query that includes only free-text terms. These methods promise to automatically identify highly effective MeSH terms for inclusion in a systematic review query. Our study contributes an empirical evaluation of several MeSH term suggestion methods. We perform an extensive analysis of the retrieval, ranking, and refinement of MeSH term suggestions for each method and how these suggestions impact the effectiveness of Boolean queries.

\end{abstract}

%
%


\maketitle

\section{Introduction \& Related Work}

A medical systematic review is a comprehensive review of literature for a highly focused research question. Systematic reviews are seen as the highest form of evidence and are used extensively in healthcare decision making and clinical medical practice. In order to synthesise literature into a systematic review, a search must be undertaken. A major component of this search is a Boolean query. The Boolean query is often developed by a trained expert (i.e., an information specialist), who works closely with the research team to develop the search, and usually has knowledge about the domain. 

The most commonly used database for searching medical literature is PubMed. Due to the increasing size and scope of the PubMed database, the Medical Subject Headings (MeSH) ontology was developed to conceptually index studies~\cite{zieman1997conceptual, richter2012using}. MeSH is a controlled vocabulary thesaurus arranged in a hierarchical tree structure (specificity increases with depth in a parent$\rightarrow$child relationship, e.g., \texttt{Anatomy}$\rightarrow$\texttt{Body Regions}$\rightarrow$\texttt{Head}$\rightarrow$\texttt{Eye}\dots etc.). Indexing and categorising studies with MeSH terms enables queries to be developed which incorporate both free-text keywords \textit{and} MeSH terms --- enabling more effective searches.
The use of MeSH terms in queries has been shown to be more effective than free-text keywords alone~\cite{chang2006searching, abdou2008searching, tenopir1985full, richter2012using}, e.g, they increase precision~\cite{liu2017evaluating} and are far less ambiguous than free-text~\cite{wacholder1997disambiguation}. 
However, it is still difficult even for expert information specialists to be familiar with the entire MeSH controlled vocabulary~\cite{liu2009impact, liu2017evaluating} --- at the time of writing, MeSH contains 29,640 unique headings.

One way that PubMed has attempted to overcome this difficulty is by developing a method called Automatic Term Mapping (ATM).
ATM is an automatic query expansion method which attempts to seamlessly map free-text keywords in a query to one of the three categories (index tables): MeSH, journal name or author name~\cite{nahin2003change}. 
Although ATM is applied by default for all queries issued to PubMed, it has several semantic limitations: it is inaccurate when used to expand free-text acronyms into MeSH terms~\cite{schulz2001indexing}; will produce different MeSH expansions even though synonymic free-text terms are used~\cite{adlassnig2009optimization}, and has difficulty disambiguating between MeSH terms and journal names~\cite{smith2004examination}. Despite these limitations, the use of ATM for MeSH term suggestion has been shown to increase the precision of free-text searches in the genomic domain~\cite{lu2009evaluation}. However, its use has, to the best of the authors knowledge, not been empirically evaluated in the context of improving the effectiveness of systematic review literature search queries.

Our paper introduces the task of MeSH term suggestion for Boolean queries used in systematic review literature search. We model this task within the context of an information specialist looking for MeSH terms to add to a query without MeSH terms present. In addition to new MeSH suggestion methods, we also propose a framework to evaluate the effectiveness of the suggestion of MeSH terms on an established collection of systematic review literature search queries. This paper adds to a recent stream of research that has focused on computational methods for the assisted creation~\cite{scells2020automatic,scells2020objective,scells2021comparison} or refinement~\cite{scells2018generating,scells2019www} of Boolean queries for systematic review creation.

The contributions of this paper are:

\begin{enumerate}[leftmargin=*]
	\item The introduction of the new task of suggesting MeSH terms for systematic review literature search (Boolean queries), modeled within the context of an information specialist looking for MeSH terms to add to a query without MeSH terms present.
	\item An empirical evaluation of the effectiveness of MeSH suggestion methods for this task (i.e., ranking MeSH terms for a query).
	\item An empirical evaluation of the effectiveness of Boolean queries using the suggestions made by different suggestion methods (i.e., retrieving abstracts for a query given different suggested MeSH terms).
\end{enumerate}



%

\begin{figure}[t!]
	\includegraphics[width=\linewidth]{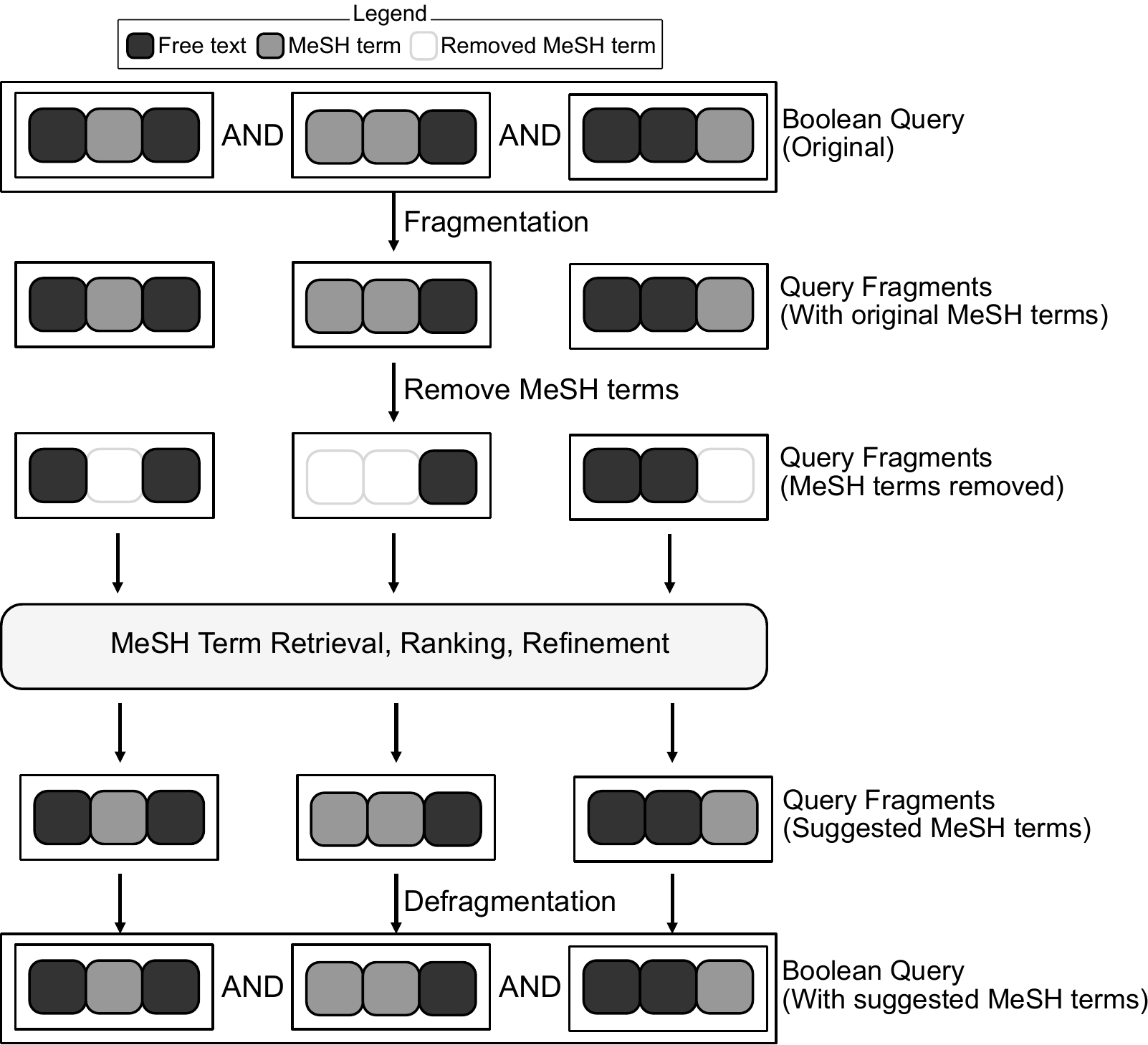}
	\caption{Overview of the MeSH term suggestion procedure. A process of retrieval, ranking and refinement facilitate the suggestion of MeSH terms. We evaluate each method that suggests MeSH terms in terms of (1) the retrieval of MeSH terms, (2) the ranking of MeSH terms, (3) the refinement of the ranking of MeSH terms, and (4) the ability for the suggested MeSH terms to effectively retrieve literature for a defragmented a Boolean query. Note that the number of MeSH terms suggested for a fragment may be lower or higher than the original number of MeSH terms.}
	\label{fig:query-overview}
\end{figure}

\section{M\texorpdfstring{\MakeLowercase{E}}{e}SH Term Suggestion}
\label{sec:methods}

Next, we outline how we perform MeSH term suggestions for Boolean queries. As the Boolean queries used for systematic review literature search are highly complex, containing nested Boolean clauses, MeSH terms are suggested not globally, but instead locally, for \textit{query fragments}. A query fragment is a clause of a Boolean query containing semantically related text clauses (i.e., free-text or \mt). Each text clause in a query fragment is grouped into a Boolean clause using the \texttt{OR} operator. To give an intuition for how query fragments are derived and utilised for MeSH term suggestions, see Figure~\ref{fig:query-overview}. The \texttt{OR} operators in Figure~\ref{fig:query-overview} are implicit. We exploit these fragments to perform a fine-grain evaluation for MeSH term suggestion (i.e., in terms of retrieval performance, ranking performance, and refinement of the ranking performance). However, we also perform defragmentation to obtain a Boolean query with suggested MeSH terms for comparison to the original Boolean queries.


We propose to suggest MeSH terms in a pipeline of three steps: retrieval, ranking, and refinement. The following three sections provide a description of how we approach each of these steps.

\subsection{MeSH Term Retrieval}

The first step in our \mts pipeline is the \textbf{retrieval} of \mt. The retrieval of \mt is facilitated by three different methods:

\begin{description}
	\item[ATM] The entire free-text only query fragment is submitted to the PubMed entrez API~\cite{sayers2010general} for ATM. When free-text clauses without specific qualifiers\footnote{Qualifiers are the terminology PubMed uses for field restrictions. Keywords in a query may be explicitly restricted to certain fields, e.g., title, abstract, \mt etc.} are present in a query, the three index tables (MeSH, journal name, author name) are searched sequentially to determine if a mapping exists. If there is no mapping found initially, the free-text clause is divided into individual terms and the process is repeated. Mapped terms are filtered to only include those that are MeSH terms.
	\item[MetaMap] Each free-text clause in a query fragment is submitted to MetaMap~\cite{aronson2001effective}.\footnote{Version 2018 with options set to default values.} The results from MetaMap are filtered to only include those entities derived from the MeSH source. All of the mapped \mt are recorded for each of the free-text terms in a query fragment. Additionally, the MetaMap score is recorded for each MeSH term.
	\item[UMLS] We index the UMLS~\cite{bodenreider2004unified} (version 2019AB) \texttt{MRCONSO}, \texttt{MRDEF}, \texttt{MRREL}, and \texttt{MRSTY} tables into Elasticsearch v7.6. Each free-text clause in the query fragment with MeSH terms removed is submitted to the Elasticsearch index. The results from the search are filtered to only include synonyms of concepts derived from the MeSH source. The synonyms of a concept are recorded for each term in the query fragment. Additionally, the BM25 score is recorded for each MeSH term (i.e., the default scoring mechanism of Elasticsearch).
\end{description}

For the MetaMap and UMLS approaches, the same MeSH term may be retrieved multiple times for a given free-text clause. To overcome this issue, we re-score the \mt using rank fusion (CombSUM)~\cite{fox1994combination}. The intuition for this re-scoring is that highly common \mt that also obtain a high score from these retrieval methods should be scored highly overall (thus ranked higher than common \mt \textit{and} highly scoring \mt).

\subsection{MeSH Term Ranking}

\begin{table}[t!]
\centering
\small
\begin{tabular}{ll}
	\toprule
	Feature & Description \\
	\midrule
$|q|$	& Total free-text terms in a fragment \\
$l_{d_e}$ & Length of description of MeSH term $e$\\
$\sum_{q_i}IDF(q_i)$ & Sum IEF of free-text terms \\
$\sum_{q_i}TF(q_i,d_e)$ & Sum TF of free-text terms in $d_e$ \\
$\sum_{q_i}TF(q_i,d_e)IDF(q_i)$ & Sum TF of free-text terms in $d_e$ \\
$\text{score}_{LM}(q,d_e)$ & LM score of free-text terms for $d_e$ \\
$\text{score}_{BM25}(q,d_e)$ & BM25 score of free-text terms for $d_e$ \\
$\text{score}_{SDM}(q,d_e)$ & SDM score of free-text terms for $d_e$ \\
$QCE(q,e)$ & Whether the free-text terms contain $e$ \\
$ECQ(q,e)$ & Whether $e$ contains any free-text terms \\
$ECQ(q,e)$ & Whether $e$ is equal to the free-text terms \\
	\bottomrule
\end{tabular}
\caption{Features used in MeSH term ranking.}
\label{table:features}
\end{table}

Once MeSH terms have been retrieved, they are ranked according to the approach for entity ranking described by Jimmy et al.~\cite{jimmy2019health} by adapting features proposed by Balog~\cite{balog2018entity}. In total, we use eleven features, each described in Table~\ref{table:features}. For the description of MeSH terms ($d_e$), we scrape the corresponding Wikipedia page. 
We generate features for each MeSH term retrieval method (i.e., ATM, MetaMap, UMLS). Positive instances correspond to \mt in the original query fragment, negative instances correspond to \mt not in the original query fragment (binary labels). 
With features and instance labels, we train a learning-to-rank (LTR) model for each MeSH term retrieval method.

In addition to the LTR models, we also investigate a rank fusion approach~\cite{fox1994combination}, where we combine the normalized MeSH term suggestion scores from each of the three methods to produce a new ranking that incorporates the highest ranking MeSH terms from each method. The intuition for investigating rank fusion in this context is that each method may retrieve different MeSH terms; and those terms may be ranked differently each time. Therefore, we wish to boost MeSH terms that are retrieved and ranked highly by each method, and further boost those MeSH terms retrieved by multiple methods.
\subsection{MeSH Term Refinement}

Finally, we seek to refine the suggested MeSH terms by estimating a rank cut-off. By refinement, we mean to limit the number of \mt to only the most applicable for a query fragment. We do this using a score-based gain function which models gain as the score for a MeSH term. Formally, the cumulative gain $CG$ for a MeSH term at rank $p$ is $CG_p = \sum_{i=1}^{p}score_i$; where the score for a MeSH term is equal to $1-normalised\ score$ (i.e., min-max normalisation) for the MeSH term. 

We tune a $\kappa$ parameter for each retrieval method which controls the percentage of total $CG$ allowed to be observed before the ranking is cut-off (i.e., a refinement of the ranking). The $\kappa$ parameter is tuned from 5\% to 95\% in increments of 5\%.
The intuition for re-scoring MeSH terms becomes apparent when used with the $\kappa$ parameter: the highest-ranking MeSH term will receive a score of 0, resulting in at least one MeSH term suggested for every query fragment. 

Note that MeSH terms may share the same score, i.e., they may be tied. We take a conservative approach to account for the problem of tied MeSH terms at the boundary of the cut-off specified by $\kappa$. Whenever we encounter ties, we treat all of the tied MeSH terms as a single accumulation of gain that equals the summed gain across the scores of the tied MeSH terms. This treatment has the effect that tied MeSH terms account for much larger accumulations of gain. Therefore, tied MeSH terms at the top of rankings are more likely to be included in the cut-off than tied MeSH terms at the bottom. In essence, either all tied MeSH terms are considered within the cut-off (i.e., ties at the top of the ranking), or no tied MeSH terms are considered (i.e., ties at the bottom of the ranking).

\subsection{Evaluation}
\label{methods:evaluation}

We evaluate the effectiveness of MeSH term suggestions retrospectively using the MeSH terms identified from pre-existing queries as a gold standard. In doing so, we make the assumption that the MeSH terms in these pre-existing queries are the ideal choices. As such, this gold standard may be biased to favour the PubMed ATM method, as it could have been used to suggest MeSH terms originally. The MeSH term suggestion methods proposed above are likely to identify MeSH terms that were not originally in pre-existing query fragments. To combat this assumption, we also evaluate the \textit{retrieval} effectiveness achieved by the queries with the proposed suggestions. We therefore evaluate both (i) the effectiveness of query suggestion given the assumption that MeSH terms in pre-existing queries are a gold standard; and (ii) the effectiveness of the query at retrieving studies. 

Note that (ii) also has limitations: that query fragments must be combined back into the original query structure in order to properly evaluate the query; and new MeSH terms may retrieve studies that are unjudged (it is unknown if the retrieved unjudged studies are relevant or not). To account for these unjudged studies, we use the approach proposed by Scells et al.~\cite{scells2019www}, which calculates, in addition to the lower bound typically assumed (i.e., all unjudged studies are irrelevant), an upper bound (i.e., assume all unjudged studies are relevant) and a balance between the two (i.e., assume some unjudged studies to be relevant given a maximum likelihood estimation over the judged studies). Note that in the paper, mle method randomly sampled unjudged studies be relevant using maximum likelihood ratio, which is equivalent to the ratio of relevant studies in the original candidate documents. 

The effectiveness of the MeSH term suggestion is evaluated using reciprocal rank, nDCG@\{5,10\}, recall@\{5,10\}, precision, and recall. Precision and recall measure the effectiveness of the retrieval of MeSH terms by the three retrieval methods. nDCG and reciprocal rank measure the effectiveness of the LTR entity ranking model for each of the three retrieval models.

To evaluate the effectiveness of the suggested MeSH terms for the task of systematic review literature search, once query fragments are defragmented, the retrieval effectiveness is evaluated using typical systematic review literature search measures: precision, recall, and F$_{\beta=\{0.5,1,3\}}$, note we will not report results for F$_{\beta=\{0.5,3\}}$ as their generally trend stay the same with F$_{\beta=\{1\}}$. To obtain retrieval results, the PubMed entrez API is used to directly issue defragmented Boolean queries. For reproducibility purposes, as PubMed is constantly updated with new studies, we apply a date restriction to all queries. 


For both evaluation settings (i.e., ranking \mts and Boolean query retrieval), we evaluate the quality of ranking in two settings: (i) \textbf{all}, where all retrieved MeSH terms are considered; and (i) \textbf{cut}, where a score-based cut-off is determined to filter the suggested MeSH terms.

\section{Experimental Setup}

We use topics from the CLEF TAR task from 2017, 2018, and 2019 \cite{kanoulas2017clef, kanoulas2018clef, kanoulas2019clef}. 15 topics are discarded due to lack of MeSH terms (\textbf{2017}: CD007427, CD010771, CD010772, CD010775, CD010783, CD010860, CD011145; \textbf{2018}: CD007427, CD009263, CD009694; \textbf{2019}: CD006715, CD007427, CD009263, CD009694, CD011768). An additional 5 topics are discarded because of retrieval issues (\textbf{2017}: CD010276, CD010173, CD012019; \textbf{2018}: CD011926; \textbf{2019}: CD010038), likely resulting from the fact that some queries are automatically translated from queries in one format (Ovid Medline) into another format (PubMed). In total we used 242 topics across all three datasets (114 unique, as each year has partial overlap). For each topic, we divide the Boolean query for that topic into several query fragments. We create these fragments using the transmute tool~\cite{scells2018querylab}. Each fragment contains at least one MeSH term. This results in a total of 302 unique query fragments for the three years (2.65 fragments per query on average). For each of the query fragments, we corrected any errors (e.g., spelling mistakes, syntactic errors), extracted MeSH terms, keywords, query fragment with MeSH terms, and query fragments without MeSH terms. 
For training the LTR model for MeSH term ranking, the pre-split training and test portions from the CLEF datasets are used. The 2019 topics are split also on systematic review type (intervention and diagnostic test accuracy --- indicated as I and D respectively in the results), while those for 2017 and 2018 are all diagnostic test accuracy. We use the quickrank library~\cite{capannini2016quality} for LTR, instantiated with LambdaMART trained to maximise nDCG. We leave other settings as per default.


%
%
%
%
%
%
%
%
\section{Results}

All of the results in this section are presented on the testing portions of each CLEF TAR year (i.e., 2017, 2018, 2019/I, 2019/D).

\begin{figure*}[t!]
	\centering
	\includegraphics[width=1\linewidth]{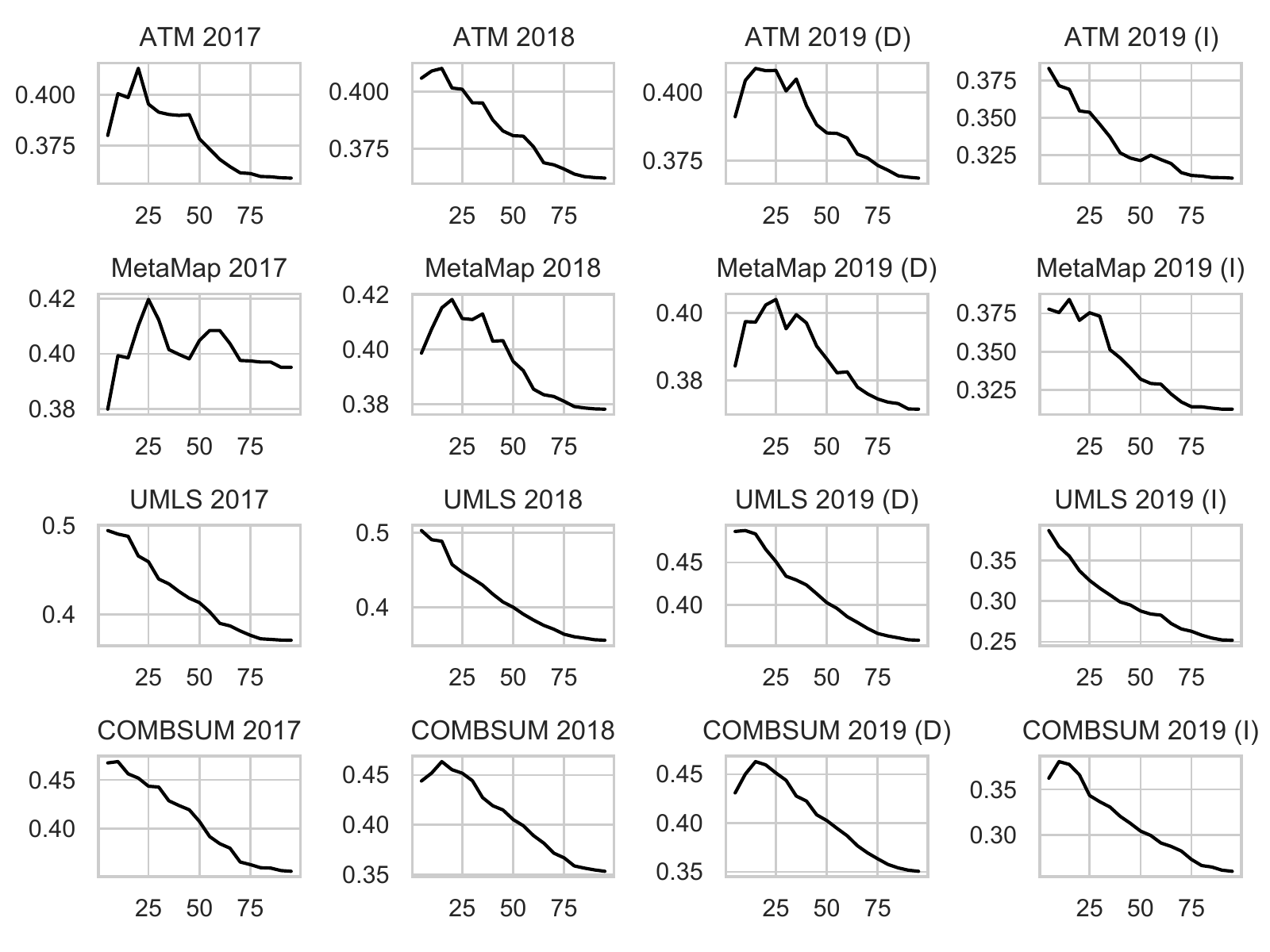}
	\vspace{-16pt}
	\caption{Tuning of the $\kappa$ parameter on training portions for each MeSH term suggestion method. The x axis is the value for $\kappa$, and the y axis is the F$-1$ at each $\kappa$ value.}
	\label{fig:tuning}
\end{figure*}

\subsection{Suggestion Effectiveness}

\begin{table}[t!]
	\centering
	\footnotesize
\begin{tabular}{p{37pt}|p{14pt}p{19pt}|p{14pt}p{14pt}p{14pt}p{26pt}p{26pt}}
	\toprule
	Method&P&R&RR&R@5&R@10&nDCG@5&nDCG@10\\\midrule
	2017/A&0.3027$^{}$&0.3718$^{}$&0.4614$^{}$&0.3504$^{}$&0.3576$^{}$&0.3601$^{}$&0.3494$^{}$\\\midrule
	2017/A-C&0.3600$^{}$&0.2421$^{*}$&0.4047$^{}$&0.2362$^{*}$&0.2403$^{*}$&0.2713$^{*}$&0.2652$^{*}$\\
	2017/M&0.3496$^{}$&0.3818$^{}$&0.5730$^{}$&0.3659$^{}$&0.3793$^{}$&0.4218$^{}$&0.4102$^{}$\\
	2017/M-C&0.4333$^{}$&0.2868$^{}$&0.4739$^{}$&0.2800$^{}$&0.2868$^{}$&0.3105$^{}$&0.3024$^{}$\\
	2017/U&0.2571$^{}$&0.4659$^{}$&0.5910$^{}$&0.4214$^{}$&0.4475$^{}$&0.4518$^{}$&0.4469$^{}$\\
	2017/U-C&\textbf{0.4819}$^{}$&0.2846$^{}$&0.5295$^{}$&0.2831$^{}$&0.2846$^{}$&0.3446$^{}$&0.3329$^{}$\\
	2017/F&0.2446$^{}$&\textbf{0.5281}$^{}$&\textbf{0.6207}$^{}$&\textbf{0.4519}$^{}$&\textbf{0.4958}$^{}$&\textbf{0.4915}$^{}$&\textbf{0.4971}$^{}$\\
	2017/F-C&0.4517$^{}$&0.3458$^{}$&0.4897$^{}$&0.3391$^{}$&0.3450$^{}$&0.3581$^{}$&0.3475$^{}$\\\midrule
	2018/A&0.3287$^{}$&0.3772$^{}$&0.4967$^{}$&0.3261$^{}$&0.3703$^{}$&0.3719$^{}$&0.3838$^{}$\\\midrule
	2018/A-C&0.3742$^{}$&0.2129$^{*}$&0.3933$^{*}$&0.1983$^{*}$&0.2024$^{*}$&0.2417$^{*}$&0.2392$^{*}$\\
	2018/M&0.3088$^{}$&0.3360$^{}$&0.4630$^{}$&0.3007$^{}$&0.3353$^{}$&0.3470$^{}$&0.3380$^{}$\\
	2018/M-C&0.3704$^{}$&0.2257$^{*}$&0.3940$^{}$&0.2237$^{}$&0.2257$^{}$&0.2689$^{}$&0.2523$^{}$\\
	2018/U&0.2885$^{}$&0.4641$^{}$&\textbf{0.6007}$^{}$&0.4188$^{}$&0.4565$^{}$&0.4615$^{}$&0.4569$^{}$\\
	2018/U-C&\textbf{0.4711}$^{}$&0.2643$^{}$&0.5278$^{}$&0.2575$^{}$&0.2643$^{}$&0.3405$^{}$&0.3292$^{}$\\
	2018/F&0.2661$^{}$&\textbf{0.5024}$^{}$&0.5793$^{}$&\textbf{0.4316}$^{}$&\textbf{0.4798}$^{}$&\textbf{0.4633}$^{}$&\textbf{0.4629}$^{}$\\
	2018/F-C&0.4212$^{}$&0.3456$^{}$&0.4754$^{}$&0.3233$^{}$&0.3387$^{}$&0.3646$^{}$&0.3550$^{}$\\\midrule
	2019/D/A&0.3399$^{}$&0.3558$^{}$&0.5933$^{}$&0.3503$^{}$&0.3558$^{}$&0.3910$^{}$&0.3671$^{}$\\\midrule
	2019/D/A-C&0.5071$^{}$&0.2628$^{}$&0.5583$^{}$&0.2628$^{}$&0.2628$^{}$&0.3222$^{}$&0.2990$^{}$\\
	2019/D/M&0.3864$^{}$&0.3053$^{}$&0.6167$^{}$&0.3003$^{}$&0.3053$^{}$&0.3810$^{}$&0.3530$^{}$\\
	2019/D/M-C&\textbf{0.5458}$^{}$&0.2303$^{}$&0.5417$^{}$&0.2253$^{}$&0.2303$^{}$&0.3209$^{}$&0.2963$^{}$\\
	2019/D/U&0.2651$^{}$&0.4528$^{}$&0.6058$^{}$&0.4428$^{}$&0.4478$^{}$&0.4680$^{}$&0.4348$^{}$\\
	2019/D/U-C&0.4850$^{}$&0.2619$^{}$&0.5167$^{}$&0.2619$^{}$&0.2619$^{}$&0.3159$^{}$&0.2866$^{}$\\
	2019/D/F&0.2266$^{}$&\textbf{0.4778}$^{}$&\textbf{0.6725}$^{}$&\textbf{0.4622}$^{}$&\textbf{0.4678}$^{}$&\textbf{0.5000}$^{*}$&\textbf{0.4680}$^{*}$\\
	2019/D/F-C&0.5068$^{*}$&0.3419$^{}$&0.4933$^{}$&0.3419$^{}$&0.3419$^{}$&0.3618$^{}$&0.3302$^{}$\\\midrule
	2019/I/A&0.3110$^{}$&0.3631$^{}$&0.4469$^{}$&0.3379$^{}$&0.3560$^{}$&0.3409$^{}$&0.3445$^{}$\\\midrule
	2019/I/A-C&0.3703$^{}$&0.2195$^{*}$&0.3868$^{}$&0.2195$^{*}$&0.2195$^{*}$&0.2511$^{*}$&0.2484$^{*}$\\
	2019/I/M&0.2651$^{}$&0.3395$^{}$&0.4253$^{}$&0.3071$^{}$&0.3368$^{}$&0.3131$^{}$&0.3205$^{}$\\
	2019/I/M-C&0.3368$^{}$&0.2178$^{}$&0.3682$^{}$&0.2088$^{}$&0.2178$^{}$&0.2389$^{}$&0.2370$^{}$\\
	2019/I/U&0.2779$^{}$&0.4175$^{}$&0.4340$^{}$&0.3765$^{}$&0.4114$^{}$&0.3516$^{}$&0.3663$^{}$\\
	2019/I/U-C&0.3415$^{}$&0.2209$^{}$&0.3769$^{}$&0.2146$^{}$&0.2209$^{}$&0.2447$^{}$&0.2464$^{}$\\
	2019/I/F&0.2566$^{}$&\textbf{0.4462}$^{}$&\textbf{0.5283}$^{}$&\textbf{0.4120}$^{}$&\textbf{0.4373}$^{}$&\textbf{0.4190}$^{}$&\textbf{0.4233}$^{}$\\
	2019/I/F-C&\textbf{0.4074}$^{}$&0.3229$^{}$&0.4336$^{}$&0.3112$^{}$&0.3166$^{}$&0.3271$^{}$&0.3255$^{}$\\\bottomrule
\end{tabular}
	\caption{Effectiveness of the MeSH term suggestion methods with respect to precision (P), recall@k (R@k), nDCG@k, and reciprocal rank (RR). \textit{A} indicates ATM, \textit{M} indicates MetaMap, \textit{U} indicates UMLS, \textit{F} indicates fusion. In each method, \textit{C} indicates cut-off ranks. Two-tailed statistical significance ( $p<0.05$) with Bonferroni correction between ATM, and the other methods, for each year is indicated by $*$.}	
\label{table:suggestion}
\end{table}

\begin{table*}
	\centering
	\small
	\begin{tabular}{p{66pt}|p{36pt}p{36pt}p{36pt}|p{36pt}p{36pt}p{36pt}|p{36pt}p{36pt}p{36pt}}
		\toprule
		Method&P&P~(MLE)&P~(Opt)&F1&F1~(MLE)&F1~(Opt)&R&R~(MLE)&R~(Opt)\\ \midrule
		2017/O&0.0298$^{}$&0.0642$^{}$&0.6962$^{}$&0.0287$^{}$&0.0893$^{}$&0.7304$^{}$&0.7490$^{}$&0.8331$^{}$&0.9035$^{}$\\
		2017/R&\textbf{0.0454}$^{}$&\textbf{0.0784}$^{}$&0.6631$^{}$&\textbf{0.0344}$^{}$&\textbf{0.0923}$^{}$&0.6675$^{}$&0.6790$^{}$&0.7396$^{}$&0.8629$^{}$\\\midrule
		2017/A&0.0276$^{}$&0.0645$^{}$&0.7963$^{*}$&0.0236$^{\dagger}$&0.0886$^{}$&0.8089$^{*\dagger}$&0.7702$^{}$&0.8338$^{}$&0.9196$^{}$\\
		2017/A-C&0.0336$^{}$&0.0690$^{}$&0.7387$^{}$&0.0280$^{\dagger}$&0.0905$^{}$&0.7529$^{\dagger}$&0.7247$^{}$&0.8035$^{}$&0.8951$^{}$\\
		2017/M&0.0319$^{}$&0.0688$^{}$&0.7866$^{}$&0.0266$^{\dagger}$&0.0904$^{}$&0.7846$^{\dagger}$&0.7668$^{}$&0.8282$^{}$&0.8980$^{}$\\
		2017/M-C&0.0359$^{}$&0.0707$^{}$&0.7450$^{}$&0.0294$^{\dagger}$&0.0908$^{}$&0.7534$^{}$&0.7221$^{}$&0.8036$^{}$&0.8902$^{}$\\
		2017/U&0.0281$^{}$&0.0658$^{}$&0.8331$^{*\dagger}$&0.0242$^{\dagger}$&0.0898$^{}$&0.8266$^{\dagger}$&0.7633$^{}$&0.8274$^{}$&0.9030$^{}$\\
		2017/U-C&0.0356$^{}$&0.0705$^{}$&0.7481$^{}$&0.0296$^{}$&0.0913$^{*}$&0.7570$^{}$&0.7346$^{}$&0.7944$^{}$&0.8906$^{}$\\
		2017/F&0.0218$^{*}$&0.0616$^{}$&\textbf{0.8577}$^{*\dagger}$&0.0194$^{*\dagger}$&0.0887$^{}$&\textbf{0.8612}$^{*\dagger}$&\textbf{0.7870}$^{}$&\textbf{0.8372}$^{}$&\textbf{0.9260}$^{}$\\
		2017/F-C&0.0340$^{}$&0.0692$^{}$&0.7621$^{}$&0.0284$^{\dagger}$&0.0905$^{}$&0.7687$^{\dagger}$&0.7263$^{}$&0.8047$^{}$&0.8943$^{}$\\
		\midrule

		2018/O&0.0215$^{}$&0.0543$^{}$&0.5721$^{}$&0.0405$^{}$&0.0971$^{}$&0.6574$^{}$&\textbf{0.8579}$^{}$&0.8892$^{}$&0.9299$^{}$\\
		2018/R&\textbf{0.0419}$^{}$&\textbf{0.0659}$^{}$&0.4290$^{}$&\textbf{0.0749}$^{}$&\textbf{0.1153}$^{}$&0.5136$^{}$&0.7827$^{}$&0.8397$^{}$&0.9064$^{}$\\\midrule
		2018/A&0.0151$^{}$&0.0542$^{}$&0.6891$^{\dagger}$&0.0277$^{}$&0.0963$^{}$&0.7484$^{\dagger}$&0.8111$^{}$&0.8794$^{}$&0.9223$^{}$\\
		2018/A-C&0.0217$^{}$&0.0568$^{}$&0.6058$^{\dagger}$&0.0403$^{}$&0.1009$^{}$&0.6681$^{}$&0.8043$^{}$&0.8669$^{}$&0.9146$^{}$\\
		2018/M&0.0180$^{}$&0.0558$^{}$&0.6481$^{\dagger}$&0.0332$^{}$&0.0993$^{}$&0.7075$^{\dagger}$&0.8020$^{}$&0.8654$^{}$&0.9162$^{}$\\
		2018/M-C&0.0197$^{}$&0.0567$^{}$&0.6031$^{}$&0.0364$^{}$&0.1008$^{}$&0.6665$^{}$&0.8009$^{}$&0.8613$^{}$&0.9154$^{}$\\
		2018/U&0.0144$^{\dagger}$&0.0542$^{}$&0.7575$^{*\dagger}$&0.0265$^{\dagger}$&0.0965$^{}$&0.8040$^{*\dagger}$&0.8173$^{}$&0.8977$^{}$&0.9254$^{}$\\
		2018/U-C&0.0287$^{}$&0.0587$^{}$&0.5786$^{\dagger}$&0.0529$^{}$&0.1042$^{}$&0.6494$^{}$&0.7892$^{}$&0.8525$^{\dagger}$&0.9130$^{}$\\
		2018/F&0.0130$^{\dagger}$&0.0540$^{}$&\textbf{0.7742}$^{*\dagger}$&0.0238$^{\dagger}$&0.0961$^{}$&\textbf{0.8192}$^{*\dagger}$&0.8359$^{}$&\textbf{0.9051}$^{}$&\textbf{0.9337}$^{}$\\
		2018/F-C&0.0168$^{}$&0.0551$^{}$&0.6697$^{\dagger}$&0.0309$^{}$&0.0980$^{}$&0.7230$^{\dagger}$&0.8066$^{}$&0.8726$^{}$&0.9165$^{\dagger}$\\
		\midrule
		
		2019/D/O&0.0199$^{}$&0.0780$^{}$&0.7179$^{}$&0.0373$^{}$&0.1286$^{}$&0.7951$^{}$&0.8966$^{}$&0.9154$^{}$&0.9871$^{}$\\
		2019/D/R&\textbf{0.0268}$^{}$&\textbf{0.0837}$^{}$&0.6850$^{}$&\textbf{0.0444}$^{}$&\textbf{0.1313}$^{}$&0.7552$^{}$&0.8326$^{}$&0.8468$^{}$&0.9408$^{}$\\\midrule
		2019/D/A&0.0096$^{}$&0.0706$^{}$&0.8546$^{}$&0.0180$^{}$&0.1159$^{}$&0.9010$^{}$&0.8916$^{}$&0.9763$^{}$&0.9984$^{}$\\
		2019/D/A-C&0.0196$^{}$&0.0782$^{}$&0.7590$^{}$&0.0342$^{}$&0.1245$^{}$&0.8134$^{}$&0.8375$^{}$&0.8524$^{}$&0.9568$^{}$\\
		2019/D/M&0.0111$^{}$&0.0715$^{}$&0.7941$^{}$&0.0209$^{}$&0.1169$^{}$&0.8554$^{}$&0.8791$^{}$&0.9340$^{}$&0.9931$^{}$\\
		2019/D/M-C&0.0172$^{}$&0.0768$^{}$&0.7494$^{}$&0.0321$^{}$&0.1260$^{}$&0.8123$^{}$&0.8393$^{}$&0.8788$^{}$&0.9806$^{}$\\
		2019/D/U&0.0097$^{}$&0.0711$^{}$&0.8421$^{}$&0.0185$^{}$&0.1166$^{}$&0.8921$^{}$&0.8616$^{}$&0.9564$^{}$&0.9977$^{}$\\
		2019/D/U-C&0.0146$^{}$&0.0745$^{}$&0.8032$^{}$&0.0262$^{}$&0.1192$^{}$&0.8488$^{}$&0.8381$^{}$&0.8713$^{}$&0.9705$^{}$\\
		2019/D/F&0.0087$^{}$&0.0707$^{}$&\textbf{0.8588}$^{}$&0.0166$^{}$&0.1160$^{}$&\textbf{0.9073}$^{}$&\textbf{0.9075}$^{}$&\textbf{0.9924}$^{}$&\textbf{0.9997}$^{}$\\
		2019/D/F-C&0.0124$^{}$&0.0723$^{}$&0.8022$^{}$&0.0231$^{}$&0.1173$^{}$&0.8572$^{}$&0.8394$^{}$&0.8841$^{}$&0.9830$^{}$\\
		\midrule
		
		2019/I/O&0.0154$^{}$&0.0637$^{}$&0.8216$^{}$&0.0204$^{}$&0.0999$^{}$&0.8085$^{}$&0.6775$^{}$&0.7397$^{}$&0.8744$^{}$\\
		2019/I/R&\textbf{0.0195}$^{}$&\textbf{0.0687}$^{}$&0.8108$^{}$&\textbf{0.0238}$^{}$&0.1018$^{}$&0.8429$^{}$&0.6381$^{}$&0.7590$^{}$&0.9146$^{}$\\\midrule
		2019/I/A&0.0140$^{}$&0.0655$^{}$&0.8599$^{}$&0.0168$^{}$&0.1031$^{}$&0.8874$^{\dagger}$&0.7268$^{}$&0.8262$^{}$&0.9375$^{}$\\
		2019/I/A-C&0.0153$^{}$&0.0661$^{}$&0.8381$^{}$&0.0174$^{}$&0.1015$^{}$&0.8680$^{\dagger}$&0.7126$^{}$&0.8150$^{}$&0.9301$^{}$\\
		2019/I/M&0.0123$^{}$&0.0629$^{}$&0.8387$^{}$&0.0200$^{}$&0.1059$^{}$&0.8852$^{}$&0.7163$^{}$&0.8126$^{}$&0.9508$^{}$\\
		2019/I/M-C&0.0145$^{}$&0.0635$^{}$&0.8193$^{}$&0.0231$^{}$&0.1044$^{}$&0.8671$^{}$&0.7012$^{}$&0.7943$^{}$&0.9390$^{}$\\
		2019/I/U&0.0107$^{}$&0.0624$^{}$&0.8622$^{}$&0.0168$^{}$&\textbf{0.1061}$^{}$&0.9007$^{}$&0.7188$^{}$&0.8232$^{}$&0.9542$^{}$\\
		2019/I/U-C&0.0153$^{}$&0.0663$^{}$&0.8404$^{}$&0.0174$^{}$&0.1023$^{}$&0.8687$^{}$&0.6808$^{}$&0.7915$^{}$&0.9266$^{}$\\
		2019/I/F&0.0098$^{}$&0.0619$^{}$&\textbf{0.8695}$^{}$&0.0157$^{}$&0.1058$^{}$&\textbf{0.9071}$^{\dagger}$&\textbf{0.7414}$^{}$&\textbf{0.8483}$^{}$&\textbf{0.9601}$^{}$\\
		2019/I/F-C&0.0142$^{}$&0.0657$^{}$&0.8406$^{}$&0.0157$^{}$&0.1023$^{}$&0.8709$^{}$&0.6976$^{}$&0.8049$^{}$&0.9304$^{}$\\
			\bottomrule
	\end{tabular}

	\caption{Effectiveness of the MeSH term suggestion when used in a Boolean query to search literature for systematic reviews. \textit{A} indicates ATM, \textit{M} indicates MetaMap, \textit{U} indicates UMLS, \textit{F} indicates fusion. In each method, \textit{C} indicates cut-off ranks. For evaluation measures, \textit{Opt} indicates optimistic treatment of residuals, \textit{MLE} indicates maximum likelihood estimation treatment of residuals. Two-tailed statistical significance ( $p<0.05$) with Bonferroni correction between the ORIGINAL query for each year and queries with new MeSH suggestions is indicated by $*$, statistical significance ( $p<0.05$) with Bonferroni correction between the Mesh term Removed query for each year and queries with new MeSH suggestions is indicated by $\dagger$.}	
	\label{table:searcglower}
\end{table*}

\subsubsection{Retrieval of MeSH Terms}
Firstly, we investigate the MeSH term suggestion methods' effectiveness in retrieving terms given a query fragment. Table~\ref{table:suggestion} reports precision (P) and recall (R) for the retrieval of MeSH terms. When comparing the three base retrieval methods, UMLS generally retrieves more relevant terms than ATM and MetaMap, as suggested by the higher recall value for UMLS than the other two methods. However, the UMLS method achieves lower precision than the other two methods, indicating that it retrieves too many MeSH terms. The fusion method achieves the highest recall across all datasets. However, it never outperforms the other methods in terms of precision (naturally because it combines all the MeSH suggestions). We find that: (i) UMLS is the most effective MeSH retrieval method for recall, (ii) ATM is the most effective retrieval method for precision, and (iii) that fusion of multiple MeSH retrieval methods generally leads to the highest recall and lowest precision.

\subsubsection{Ranking of MeSH Terms}

Next, we investigate the effectiveness of the LTR model at ranking the retrieved MeSH terms for each retrieval method. For this task, we observe the reciprocal rank (RR), R@k (Recall@k), and nDCG@k of the results reported in Table~\ref{table:suggestion}. We find that (i) because the UMLS method generally has the highest recall compared to ATM and MetaMap, the ranking performance was also generally higher than these methods in most measures and (ii) also, due to the higher recall, the fusion method produces more effective rankings of MeSH terms, except for RR on the 2018 dataset. 

\subsubsection{Refinement of MeSH Terms}

Finally, we investigate the effect of refining the ranked MeSH terms by cutting off the ranking at a certain point and discarding the remainder. We estimate this cut-off point through a parameter. Our tuning results on the training portions of the datasets are presented in Figure~\ref{fig:tuning}. We believe that the spikes in these plots generally correspond to the inclusion and exclusion of ties. These spikes are most prominent in the MetaMap, and ATM methods as these methods do not assign highly discriminative scores to MeSH terms. Furthermore, note that the UMLS and fusion methods have considerably smoother shapes, as these methods have highly discriminative scores.

We investigate the effect that this refinement has on the MeSH term suggestion performance in Table~\ref{table:suggestion} (i.e., with -C). We find that refinement generally improves precision while lowering recall. The loss in recall attributed to the refinement negatively affects ranking effectiveness. In most cases, refinement is worse than ranking all of the MeSH terms and often significantly worse than the ATM baseline (without refinement).


\subsection{Search Effectiveness}
\label{sec:results.effectiveness}

\begin{table*}
	\centering
	\footnotesize
	\begin{tabular}{l|p{145pt}|c|p{195pt}|l|p{75pt}}
		\toprule
		& \multicolumn{0}{c}{Fragment 1} & \multicolumn{0}{c}{} &\multicolumn{1}{c}{Fragment 2} &\multicolumn{0}{c}{} &\multicolumn{1}{c}{Fragment 3} \\\midrule
		O & \textbf{Elasticity Imaging Techniques} OR transient elastograph OR fibroscan                                     &                         & \textbf{liver cirrhosis} OR (hepatic OR liver) AND (fibrosis OR cirrhosis)                                                                                                        &                         &      liver biops OR \textbf{Biopsy, Needle}      \\
		\cmidrule{1-2}\cmidrule{4-4}\cmidrule{6-6}
		R  & transient elastograph OR fibroscan                                                                               &                         & (hepatic OR liver) AND (fibrosis OR cirrhosis)                                                                                                                                    &                         &                   liver biops                    \\ \cmidrule{1-2}\cmidrule{4-4}\cmidrule{6-6}
		A      & \textbf{transients and migrants} OR \textbf{elasticity imaging techniques} OR transient elastograph OR fibroscan & \multirow{5}{12pt}{ AND} & \textbf{liver} OR \textbf{fibrosis} OR \textbf{liver cirrhosis} OR (hepatic OR liver) AND (fibrosis OR cirrhosis)                                                                 & \multirow{5}{12pt}{AND}                        & \textbf{liver} OR \textbf{biopsy} OR liver biops \\ \cmidrule{1-2}\cmidrule{4-4}\cmidrule{6-6}
		M  & transient elastograph OR fibroscan                                                                               && \textbf{liver} OR \textbf{fibrosis} OR \textbf{liver cirrhosis} OR (hepatic OR liver) AND (fibrosis OR cirrhosis)                                                                 &  & \textbf{liver} OR \textbf{biopsy} OR liver biops \\ \cmidrule{1-2}\cmidrule{4-4}\cmidrule{6-6}
		U     & transient elastograph OR fibroscan                                                                               &                         & \textbf{fibrosis} OR \textbf{hepatic artery} OR \textbf{liver} OR \textbf{liver cirrhosis} OR \textbf{genetic diseases, inborn} OR (hepatic OR liver) AND (fibrosis OR cirrhosis) &                         &          \textbf{liver} OR liver biops           \\\bottomrule	\end{tabular}
	\caption{Query fragments in different methods,  \textit{O} indicates original query, \textit{R} indicates MeSH term removed query,  \textit{A} indicates ATM, \textit{M} indicates MetaMap, \textit{U} indicates UMLS. In each method. For evaluation measures, bold text means mesh term.}	
	\label{table:casestudy}
\end{table*}

We next investigate the impact in performance that MeSH term retrieval, ranking, and ranking refinement has on the retrieval effectiveness of Boolean queries.
\subsubsection{Impact of MeSH Terms}
Comparing the original query to the original query with MeSH terms removed, a general trend in every dataset is that the removal of MeSH terms results in a tradeoff where precision increases and recall decreases. Comparing the original query with the MeSH term suggestion methods, the same trend also appears, which indicates that the addition of relevant MeSH terms will increase the number of relevant studies. 

The retrieval of literature for systematic reviews is a high recall task. Our MeSH term suggestion methods can automatically support this high recall task by recommending appropriate MeSH terms given a Boolean query without MeSH terms. The results obtained from many of the MeSH term suggestion methods presented in this work are comparable to those achieved by queries initially constructed by information specialists.

\subsubsection{Impact of Unjudged Studies}
We next examine the retrieval effectiveness when we consider unjudged studies to be irrelevant. This assumption is a typical retrieval evaluation scenario and provides a lower bound on effectiveness. For the 2017 and 2018 datasets, we find that few methods increase precision over the original queries (both are refined rankings); however, for the two 2019 datasets, there is no method where we see an increase in precision over the original queries. However, none of the results obtained statistically significantly worse results except for 2017/F.

Comparing these results to our optimistic and MLE residual treatments of unjudged studies, we find that (i) the unrefined fusion ranking achieves the highest results in recall across all datasets, likely a result of the fact that it retrieves the most MeSH terms; (ii) Although we find that unrefined fusion still achieves the highest recall for the MLE treatment, it generally performs worse than other methods.

\subsubsection{Impact of Fusion}
\label{Impact of Fusion}
In terms of recall, the unrefined fusion ranking improved recall except for a single case (2018). This large gain in recall is likely because unrefined fusion combines all of the MeSH terms suggested by the other three methods (ATM, MetaMap, and UMLS).  This suggests that the unrefined fusion method is not beneficial for improving the precision of a Boolean query. However, suppose semi-automatic MeSH term suggestion can be used. Information specialists may be able to use the suggestion and apply their expertise to decide which MeSH terms should be included to achieve higher performance.

To our surprise, the refined fusion method did not achieve the highest result among any evaluation measure or dataset. Indeed, refinement of rankings generally lowered recall and had a negligible effect on precision for all methods. This suggests that choosing appropriate MeSH terms is crucial for effective systematic review literature retrieval instead of adding as many MeSH terms as possible. 

\vspace{-4pt}

\subsection{Case Study}
Based on the findings in Section~\ref{sec:results.effectiveness}, it may also be interesting to observe a concrete example of MeSH term suggestion and why the effectiveness may vary for different suggestion methods. After exploring different queries produced from our MeSH suggestion methods, we chose CD010542 from the 2017 CLEF TAR dataset to conduct our analysis because it is a representative example. We follow the same procedure as in Figure~\ref{fig:query-overview} by removing all MeSH terms, fragmenting the query into separate clauses, suggesting MeSH terms for each clause, and defragmenting the clauses with suggested MeSH into a Boolean query we use for retrieval. We use the re-constructed queries to search PubMed and compare their effectiveness with the original query and the query without any MeSH terms. Query fragments in the procedure and suggested method are shown in Table~\ref{table:casestudy}.

In Table~\ref{table:case}, we found that the scores obtained using ATM and Metamap are the same while their first fragment is different. This strengthens our hypothesis from Section~\ref{Impact of Fusion} that the choice of MeSH terms is crucial to the effectiveness of a Boolean query in systematic review literature search.  When comparing UMLS with other methods, we found that UMLS obtained a higher precision when compare with other query with MeSH terms except when optimistic measurements were used, this suggests that the MeSH term from fragment 1 in the original query may be detrimental to the effectiveness of the query; therefore, even the MeSH terms in the original queries may not be the most effective choice.

Finally, the query with MeSH terms removed achieved a higher precision than any other methods and maintained the same recall value, this suggests that MeSH terms may not be effective to some queries, a more dynamic method to add the most appropriate MeSH terms may have the potential to further improve performance of MeSH term suggestion. We propose that: (i) For a fully automatic pipeline, a classification model can be trained and used to decide the effectiveness gain of a MeSH term, a stopping strategy can also be used to make the best decision of when to stop adding terms to obtain the best performance.  (ii) For a semi-automatic pipeline, the classification model can be reused to compute MeSH terms' confidence scores recurrently, this may provide a better understanding for information specialists to decide on which MeSH terms to use when constructing the new query.

\begin{table}[t!]
	\centering
	\footnotesize
	\begin{tabular}{p{18pt}|p{10pt}p{16pt}p{16pt}|p{10pt}p{17pt}p{18pt}|p{10pt}p{16pt}p{16pt}}
		\toprule
		Method&P&P~(MLE)&P~(Opt)&F1&F1~(MLE)&F1~(Opt)&R&R~(MLE)&R~(Opt)\\ \midrule
		O&0.0207&0.0598&0.6622&0.0405&0.1126&0.7968&0.9000&1.0000&1.0000\\\midrule
		R&0.0274&0.0608&0.6140&0.0531&0.1143&0.7594&0.9000&0.9524&0.9951\\\midrule
		A&0.0167&0.0583&0.7574&0.0327&0.1100&0.8611&0.9000&0.9692&0.9976\\ \midrule
		M&0.0167&0.0583&0.7574&0.0327&0.1100&0.8611&0.9000&0.9692&0.9976\\ \midrule
		U&0.0256&0.0598&0.6382&0.0499&0.1126&0.7778&0.9000&0.9545&0.9956\\
		
		\bottomrule
	\end{tabular}
	\caption{Effectiveness of the MeSH term suggestion when used in a Boolean query to search literature for systematic reviews for CD010542, \textit{O} indicates original query, \textit{R} indicates MeSH term removed query, \textit{A} indicates ATM, \textit{M} indicates MetaMap, \textit{U} indicates UMLS.}	
	\label{table:case}
\end{table}

\section{Conclusions}

In this paper, we presented the new task of suggesting MeSH terms within the context of systematic review literature search (suggestion for Boolean queries). We provided a comprehensive evaluation of the effectiveness of MeSH suggestion methods (in terms of retrieval, ranking, and refinement). We compared these methods to the existing method that PubMed uses to suggest MeSH terms (ATM). We found that both the MetaMap and UMLS suggestion methods can improve the retrieval effectiveness of Boolean queries. Unsurprisingly, when we combined the three methods using rank fusion, we found the highest gains in retrieval effectiveness. 

Our methods overcome the semantic limitations of ATM: the MetaMap and UMLS methods both suggested more relevant MeSH terms than ATM, and the addition of these terms positively impacted retrieval performance. Often this came with a minor loss in recall. Note that there are generally between 10-100 relevant studies per topic: the actual impact of loss in recall is attributed to only a handful of studies and is likely not to impact the results of a systematic review.

Identifying MeSH terms to add to a Boolean query for systematic review literature search is known to be a difficult task for humans to accomplish. The outcomes of this paper have implications for both the information retrieval and systematic review communities. Firstly, our methods can be used in automatic query formulation situations (see, e.g., tasks in CLEF TAR). Secondly, they can be integrated into existing tools to assist information specialists in formulating more effective queries~\cite{scells2018searchrefiner,li2020systematic}. 

\subsubsection*{Acknowledgement} 
This research is supported by the Australian Research Council (DP210104043).
Dr Guido Zuccon is the recipient of an Australian Research Council DECRA Research Fellowship (DE180101579). 

\bibliographystyle{ACM-Reference-Format}
\bibliography{references.bib}

\end{document}